\documentclass[aps,pre,amsfonts,amssymb,amsmath,floatfix,superscriptaddress,reprint,longbibliography]{revtex4-1}

\usepackage{times}
\usepackage{epsfig}
\usepackage{graphicx}
\usepackage{epstopdf}
\usepackage{bm}
\usepackage{color}
\usepackage[english]{babel}
\usepackage{microtype}

\PassOptionsToPackage{hyphens}{url}
\usepackage[breaklinks=true,colorlinks=true,citecolor=blue,urlcolor=blue,linkcolor=blue]{hyperref}

\renewcommand{\vec}[1]{\mathbf{#1}}

\newcommand{\pd}{\partial}

\begin{document}

\title{Kinetic Turbulence in Astrophysical Plasmas: Waves and/or Structures?}

\author{Daniel Gro\v selj}
\affiliation{Max-Planck-Institut f\" ur Plasmaphysik, D-85748 Garching, Germany}

\author{Christopher H. K. Chen}
\affiliation{School of Physics and Astronomy, Queen Mary University of London, London E1 4NS, UK}

\author{Alfred Mallet}
\affiliation{Space Sciences Laboratory, University of California, Berkeley CA 94720, USA}
\affiliation{Space Science Center, University of New Hampshire, Durham, NH 03824, USA}

\author{Ravi Samtaney}
\affiliation{Mechanical Engineering, 
King Abdullah University of Science and Technology, Thuwal 23955-6900, Saudi Arabia}

\author{Kai Schneider}
\affiliation{I2M-CNRS, Centre de Math\' ematiques et d'Informatique, Aix-Marseille Universit\' e, 13453 Marseille Cedex 13, France}

\author{Frank Jenko}
\affiliation{Max-Planck-Institut f\" ur Plasmaphysik, D-85748 Garching, Germany}

\date{\today}

\begin{abstract}

The question of the relative importance of coherent structures and waves 
has for a long time attracted a great deal of interest in astrophysical plasma turbulence research, with a 
more recent focus on kinetic scale dynamics. Here we utilize 
high-resolution observational and simulation data to investigate the nature of waves and 
structures emerging in a weakly collisional, turbulent kinetic plasma. Observational results are based on \emph{in situ} 
solar wind measurements from the Cluster and MMS 
spacecraft, and the simulation results are obtained from an externally driven, three-dimensional fully kinetic simulation. 
Using a set of novel diagnostic measures we show that both the large-amplitude structures 
and the lower-amplitude background fluctuations preserve linear features of kinetic Alfv\' en waves
to order unity. This quantitative evidence suggests that the kinetic turbulence cannot be described as a mixture of mutually 
exclusive waves and structures but may instead be 
pictured as an ensemble of localized, anisotropic wave packets or ``eddies'' of varying amplitudes, which preserve certain linear wave 
properties during their nonlinear evolution.
\end{abstract}

\maketitle

\section{INTRODUCTION}

A wide range of turbulent systems, in different areas of physics, supports linear waves 
\citep{Tu1995, Lindborg2006, Denissenko2007, Proment2009, Nazarenko2011a, Yokoyama2014, Godeferd2015}. Prominent 
examples are represented by magnetized plasmas \citep{Tu1995,Krommes2002,Chen2016}, 
rotating and/or stably stratified fluids \citep{Lindborg2006,Godeferd2015}. Waves in 
turbulent environments are typically supported by a mean field, such as a mean magnetic field in a plasma or a mean 
rotation and gravity field in geophysical flows \citep{Davidson2013}. 
Not only does the mean field introduce waves, it also renders the 
turbulence inherently anisotropic and sets the stage for a competition between linear and nonlinear dynamics, 
the outcome of which is presently the subject of ongoing debates 
\citep{Lindborg2006,Staplehurst2008,Nazarenko2011,ClarkdiLeoni2014,Matthaeus2014,Howes2015a,Meyrand2016}.
Alongside wavelike phenomena, turbulence exhibits striking features of 
self-organization in the form of coherent structures, such as electric current sheets in magnetized plasmas or
columnar vortices in rotating fluids \citep{Benkadda1994,Smolyakov2000,Sundkvist2005,
Staplehurst2008,Zhdankin2013,Matthaeus2015,
Biferale2016,Perrone2016,Mordant2017,Camporeale2018}. Although these intense structures
are believed to significantly impact the turbulence statistics and the 
rate of energy dissipation \citep{Sundkvist2007,Salem2009,Wu2013a,Matthaeus2015,
Biferale2016,Camporeale2018}, their relation to wave dynamics remains poorly understood.

A series of compelling questions
emerge in the above context such as: Which of the two aspects, structures or waves, 
is more essential and/or fundamental to the turbulent dynamics? 
And also: Are waves and structures mostly exclusive to each other and simply ``coexist'' in the turbulence or is there a 
deeper connection between the two?
Here, we address these open questions in the context of 
kinetic-scale plasma turbulence, where the topic of 
waves and structures is hotly debated at present \citep{Sundkvist2007,Salem2012,TenBarge2013,Wu2013a,Chen2013,
Osman2014,Wan2015,Howes2015a,Lion2016,Perrone2016,Klein2017,Groselj2018,Camporeale2018}.
In the following, we explain the astrophysical background and motivation for the present study.  
Our general approach and the qualitative conclusions drawn from it are, however, not exclusively limited to 
kinetic-scale plasma turbulence and could be relevant to a broad range of turbulent 
regimes and environments where waves and structures can be identified.

\subsection{Astrophysical background and motivation}

In magnetized astrophysical and space plasmas, the role of waves in turbulence 
has been widely debated since the early days of magnetohydrodynamic (MHD) turbulence research when 
first models based on nonlinearly interacting Alfv\' en waves were proposed \citep{Iroshnikov1963,Kraichnan1965}.
Soon after, \emph{in situ} spacecraft 
measurements detected Alfv\' en wave 
signatures in the turbulent solar wind \citep{Belcher1971}.
Observations and models were followed by pioneering 
numerical studies (e.g., \citep{Orszag1979, Matthaeus1980, Meneguzzi1981}).
These simulations revealed the development 
of intense electric current sheets, which are now believed to play a key 
role in the (Ohmic) dissipation of MHD turbulence \citep{Uritsky2010,Zhdankin2013,Matthaeus2015}. 
It also soon became clear that magnetized plasma 
turbulence is inherently anisotropic, as evidenced by the
preferential cascade of energy to small perpendicular scales with respect to 
the direction of the mean magnetic field \citep{Shebalin1983,Matthaeus1990,Oughton1994}. 

Motivated by the observational and numerical evidence available at the time, 
Goldreich \& Sridhar \citep{Goldreich1995} devised their celebrated model 
of anisotropic, strong MHD turbulence based on the \emph{critical balance} conjecture. 
In short, they assumed a balance between the linear (wave-crossing) 
and nonlinear (eddy turnover) time at each scale. This allows for a 
phenomenological prediction of the turbulence energy spectrum and its
anisotropy. The latter is presently consistent with 
numerical simulations \citep{Cho2000,Maron2001,Mallet2015} and recent 
observations \citep{Horbury2008,Podesta2009,Wicks2010,Chen2016}. Although the Goldreich \& Sridhar 
model is by no means a rigorous description of MHD-scale turbulence, their 
phenomenological approach received a great deal of interest in the community and it later paved the way for 
many notable works on astrophysical plasma turbulence 
(e.g., \citep{Lazarian1999,Boldyrev2006,Howes2008,Schekochihin2009}). 

It is worth noticing that the critical balance conjecture \citep{Goldreich1995} 
is essentially a statement about the persistence of 
linear wave physics in a strongly turbulent wave system. The rather generic nature of
the statement implies that critical balance
may be relevant in a more general context outside the scope of MHD turbulence.
Indeed, the concept has been more recently 
adapted in models of sub-ion scale plasma 
turbulence \citep{Howes2008,Schekochihin2009}, and it even received 
attention in fields outside of plasma 
physics \citep{Proment2009,Nazarenko2011,Yokoyama2014}. In particular, 
Ref.~\citep{Nazarenko2011} proposed critical balance as a 
universal scaling conjecture for strong turbulence in wave systems, with application to
MHD, rotating, and stratified flows. 

With advances in observational and computational techniques, focus 
has shifted over the years from 
MHD towards kinetic turbulence at scales below the ion thermal gyroradius, 
where one of the ultimate goals is to understand how weakly collisional, 
turbulent astrophysical plasmas dissipate energy and heat the ions and
electrons \citep{Quataert1999,Cranmer2003,Sundkvist2007,Wu2013a,TenBarge2013a,TenBarge2013,
Zhuravleva2014,Wan2015,BanonNavarro2016,Klein2017,Kawazura2019}. Research in the area has 
been largely driven by \emph{in situ} solar wind 
observations \citep{Chen2016}, with the hope that the key findings could be 
extrapolated to other astrophysical systems of interest such 
as galaxy clusters \citep{Dennis2005,Zhuravleva2014}, hot accretion 
flows \citep{Quataert1999,Kunz2016,Kawazura2019}, and the warm interstellar 
medium \citep{Gaensler2011, Haverkorn2013}. Early observations 
(e.g., \citep{Leamon1998,Bale2005,Sahraoui2009,Alexandrova2009})
were soon complemented by theory. About a decade ago, a well known phenomenological model of kinetic range 
turbulence was introduced in the works of Schekochihin \emph{et al.}~\citep{Schekochihin2009} 
and Howes \emph{et al.}~\citep{Howes2008}. 
Their model, commonly referred to as kinetic Alfv\' en wave (KAW) turbulence, follows the idea of 
Goldreich \& Sridhar and 
assumes critical balance between the nonlinear and linear (KAW) time. 
A large body of observational and numerical evidence in support of the KAW turbulence 
phenomenology has been presented in recent years (see Sec.~\ref{sec:objectives}), although
concerns have been raised as well (e.g., \citep{Matthaeus2008,Podesta2010}) and 
the subject remains open. In terms of wavelike interpretations, the main 
alternative to KAW turbulence is the whistler turbulence model \citep{Biskamp1999,Galtier2003,
Cho2004,Gary2009,Shaikh2009,Gary2012}.
Complementary to wavelike models of kinetic turbulence, several authors emphasized 
the role of intermittency and the nonlinear generation of turbulent structures, such as current sheets 
\citep{Sundkvist2007,Karimabadi2013a,Osman2014,Chasapis2015,Phan2018}, 
Alfv\' en vortices~\citep{Sundkvist2005,Lion2016,Perrone2016,Wang2019}, 
or magnetic holes~\citep{Haynes2015,Roytershteyn2015,Huang2017}.  
Indeed, \emph{in situ} solar wind measurements~\citep{Sundkvist2007,Roberts2013,Osman2014,Lion2016, 
Chasapis2015, Perrone2016} and numerical simulations~\citep{Boldyrev2012,
Karimabadi2013a, Servidio2015, Wan2015, Haynes2015, Yang2017, Kobayashi2017,Camporeale2018} 
support the idea that kinetic-scale coherent structures emerge naturally as a result 
of the turbulent cascade. 

The discussion on waves versus structures became 
more pronounced in the kinetic plasma turbulence context, since the two viewpoints lead to fundamentally different 
interpretations of how turbulent dissipation works. In a wave-mediated type of
turbulence, there is hope that at least a significant fraction of heating can be estimated from 
linear (collisionless) wave damping rates \citep{Howes2011,TenBarge2013a,
BanonNavarro2016,Kobayashi2017,Howes2018,Chen2019}, whereas a 
structure-mediated regime might first require an understanding of how
structures form and dissipate energy during their evolution \citep{Wu2013a,Karimabadi2013a,Osman2014,Wan2015,Chasapis2015,Camporeale2018}. 
It is important to notice that the two viewpoints are not \emph{a priori} incompatible, 
if the kinetic-scale structures themselves were to preserve certain wavelike features. The latter 
possibility is explored in the present paper.

\subsection{Our approach and relation to previous works}
\label{sec:objectives}

We employ the KAW turbulence 
phenomenology~\citep{Howes2008,Schekochihin2009,Boldyrev2013} as the main basis for making contact with linear wave 
predictions. 
It is not \emph{a priori} obvious that the sub-ion scale fluctuations are of the KAW type. Other
possibilities, such as whistler waves \citep{Galtier2003,Gary2009,Shaikh2009}, 
kinetic slow waves \citep{Narita2015}, or ion Bernstein modes \citep{Podesta2012} have been considered as well.
Our choice is, however, favored by the presently available empirical evidence 
(see Podesta \citep{Podesta2013} for a detailed review). A significant portion of evidence is based on the measured 
statistical polarization properties of the turbulent fields 
\citep{Bale2005,Sahraoui2009,Podesta2011,Salem2012,He2012,TenBarge2012a,Chen2013,
Kiyani2013,Klein2014,Cerri2017,Groselj2018}, which have been compared against linear KAW predictions.
This includes, for instance, the measured spectral ratio between the parallel and perpendicular 
magnetic field component (e.g., \citep{Salem2012,TenBarge2012a,Kiyani2013}), and 
the spectral ratio between the electron density and the magnetic field (e.g., \citep{Chen2013}).
Some works considered as well the magnetic helicity and related its observed spectral signatures
to the right-hand elliptical polarization of the KAW \citep{Podesta2011,He2012,Klein2014}.
Apart from the above, a number of computational and observational studies suggest that the sub-ion scale 
fluctuations are predominantly low-frequency \citep{Sahraoui2010,Kobayashi2017} 
and strongly anisotropic \citep{Chen2010,Sulem2016,Chen2017,Groselj2018}, as 
qualitatively expected for KAW turbulence \citep{Howes2008,Schekochihin2009,Boldyrev2012,Boldyrev2013}.

Previous works supporting wavelike interpretations of kinetic-scale plasma turbulence were 
often focused on turbulence spectral features, while paying relatively little attention to 
the characteristic structure of the fluctuations in real space and 
their intermittent statistics \citep{Karimabadi2013a,Wu2013a,Leonardis2013,Chen2014,
Servidio2015,Wan2015,Matthaeus2015,Perrone2016}.
In order to account for turbulent structure formation it is necessary to go beyond 
a standard spectral analysis. To this end, we introduce here appropriate generalizations of the 
frequently employed spectral field ratios. Unlike the standard field ratios, 
the generalized definitions are specifically designed to probe the statistical field 
polarizations within the large-amplitude turbulent structures. This is accomplished 
with the aid of a wavelet scale decomposition 
\citep{Torrence1998, Kiyani2013,Farge2015, Perrone2016, Lion2016}, 
which can be used to analyze a signal simultaneously in real space and spectral space.

Our newly defined diagnostic measures are applied jointly to high-resolution observational and 
kinetic simulation data. Observational results are based on \emph{in situ} measurements from the 
Cluster and MMS spacecraft, whereas the numerical results are obtained from an externally driven, 
three-dimensional (3D) fully kinetic particle simulation. 
The majority of previous kinetic 
simulations were carried out either in a two-dimensional (2D)
geometry and/or with a reduced-kinetic model. The validity of both types of 
approximations has been questioned \citep{Matthaeus2008,Boldyrev2013,Howes2015,
Servidio2015,Told2016,Groselj2017}. Specifically for KAW turbulence,
pioneering 3D kinetic simulations were carried out using the gyrokinetic formalism
\citep{Howes2008a,Howes2011,TenBarge2013,TenBarge2013a}. These works 
obtained results consistent with theoretical expectations. However, gyrokinetics assumes 
low-frequency (compared to the ion cyclotron motion), strongly anisotropic (with respect
to the mean field), and small-amplitude fluctuations \citep{Howes2006,Schekochihin2009}, which
naturally favors a KAW-type of turbulent cascade. 
Only recently, various aspects of KAW turbulence were studied 
in terms of 3D hybrid-kinetic (i.e., with fully kinetic ions and fluid electrons) 
and fully kinetic simulations \citep{Cerri2017,Hughes2017,Franci2018,Groselj2018,Arzamasskiy2019}.

The rest of this paper is organized as follows. In Sec.~\ref{sec:theory} we 
provide a short theoretical background. 
The methods employed are described in Sec.~\ref{sec:methods}. This includes a brief overview 
of our 3D fully kinetic simulation, a description of observational data, a summary 
of wavelet scale decompositions, and finally, the generalized field ratio definitions. 
Our main results are presented in Sec.~\ref{sec:results}. We conclude the paper 
with a discussion (Sec.~\ref{sec:discussion}) 
and summary (Sec.~\ref{sec:conclusions}) of our results.

\section{THEORETICAL BACKGROUND}
\label{sec:theory}

The essential plasma dynamics underlying kinetic-Alfv\' en turbulence may be approximately
described with a set of ideal (i.e., dissipation-free 
\footnote{In practice, the ideal equations are
supplemented with small dissipation terms to remove energy 
at the smallest (resolved) scales \citep{Boldyrev2012,Boldyrev2013}.}) fluid equations \citep{Schekochihin2009, Cho2009, Boldyrev2013}
 for the electron density $n_e$ and
flux function $\psi$, defined via $\vec b_\perp = \hat{\vec z}\times\nabla_\perp\psi$, where 
$\nabla_\perp = (\pd_x, \pd_y)$ and $\vec b_\perp$ is the part of the perturbed magnetic field 
perpendicular to the mean field $\vec B_0 = B_0\hat{\vec z}$.
It is convenient to first normalize the density and magnetic 
field according to \citep{Cho2009,Boldyrev2013,Chen2013}:
\begin{align}
 n_e'& = \left[\frac{\beta_i + \beta_e}{2}\left(1 
+ \frac{\beta_i + \beta_e}{2}\right)\right]^{1/2}\frac{n_e}{n_0},& \label{eq:norm_den}\\
\vec b_{\perp}'& = \frac{ \vec b_{\perp}}{B_0},& \label{eq:norm_bperp} \\
\vec b_{\parallel}'& = \left(\frac{2 + \beta_i + \beta_e}{\beta_i + \beta_e}\right)^{\!1/2}\frac{\vec b_{\parallel}}{B_0},& \label{eq:norm_bpar}
\end{align}
where $\vec b_\parallel=b_\parallel\hat{\vec z}$ 
is the parallel perturbed magnetic field, $n_0$ is the mean density, 
$\beta_i= 8\pi n_0T_i/B_0^2$ and $\beta_e=8\pi n_0T_e/B_0^2$ are the 
ion and electron betas, and $T_i$ and $T_e$ are the ion and electron temperatures (measured in energy units), respectively.
From here on, we drop the prime signs but it is to be understood that all fields 
are normalized according to \eqref{eq:norm_den}-\eqref{eq:norm_bpar} unless noted otherwise.
In appropriately chosen time and length units (see Ref.~\citep{Boldyrev2013}), the 
fluidlike equations read
\begin{align}
\pd_t\psi & = - \pd_z n_e - \hat{\vec z}\cdot(\nabla_\perp\psi\times\nabla_\perp n_e), & 
\label{eq:ermhd1}
\\
\pd_t n_e & =  \pd_z\nabla_\perp^2\psi + 
\hat{\vec z}\cdot(\nabla_\perp\psi\times\nabla_\perp\nabla_\perp^2\psi). & 
\label{eq:ermhd2}
\end{align}
Parallel magnetic fluctuations are not evolved explicitly but are 
instead determined by a linearized pressure balance condition:
 \begin{align}
n_e - n_0 = - b_{\parallel},
\label{eq:p_balance}
\end{align}
where $n_0$ is now the mean electron density in the normalized units.
Note 
that Eqs.~\eqref{eq:ermhd1}-\eqref{eq:p_balance} are obtained for a $\beta\sim 1$ plasma 
in the asymptotic limit
\begin{align}
1/\rho_i \ll k_{\perp} \ll 1/\rho_e, && k_{\parallel}\ll k_{\perp},
\label{eq:asymptotic_limit}
\end{align}
where $k_{\parallel}$ is the parallel wave number, $k_{\perp}$ is the perpendicular 
wave number, $\rho_i$ is the ion thermal gyroradius, and $\rho_e$ is the electron thermal gyroradius.

The above model is essentially nonlinear. By dropping the second term on the 
right-hand sides of \eqref{eq:ermhd1} and \eqref{eq:ermhd2}, a linear wave system is 
obtained, the non-stationary solutions of which may be written as linear combinations of plane 
KAWs with polarizations \citep{Schekochihin2009,Boldyrev2013}:
\begin{align}
 n_{e\vec k} & = \pm k_\perp \psi_{\vec k}, \label{eq:kaw_polarization0} \\
\bigl| n_{e\vec k}\bigr|^2 & =  \bigl|\vec b_{\perp\vec k}\bigr|^2 = 
\bigl|\vec b_{\parallel\vec k}\bigr|^2
=  {\mathcal E}_{{\rm KA},\vec k}/2 ,
\label{eq:kaw_polarization}
\end{align}
for each mode with a wave vector $\vec k$, where ${\mathcal E}_{\rm KA,\vec k} = \bigl|\vec b_{\perp\vec k}\bigr|^2 
+ \bigl| n_{e\vec k}\bigr|^2$ is the KAW energy density. 
Adopting the convention $k_\parallel \geq 0$, the frequency of 
the waves is $\omega_{\vec k} = \pm k_\parallel k_\perp$.
Similarly to inertial waves in rotating fluids \citep{Nazarenko2011,Davidson2013,Yarom2014}, 
KAWs are dispersive and their group velocity is preferentially oriented 
along the mean field (assuming $k_\parallel \ll k_\perp$). Besides KAWs, solutions of the
linear system are also static structures with $k_\parallel=0$.

Following previous works (e.g., \citep{Salem2012,TenBarge2012a,Chen2013,Groselj2018}), we use
relation \eqref{eq:kaw_polarization} as the basis for identifying 
kinetic-Alfv\' en wave features in 
a nonlinear, strongly turbulent regime \footnote{Following \citet{Boldyrev2013}, waves corresponding to the extension of 
the kinetic-Alfv\' en branch above the ion cyclotron frequency, sometimes referred to as Alfv\' en-whistler 
modes \citep{Sahraoui2012,Gershman2018}, would be identified as KAWs rather than 
whistlers according to \eqref{eq:kaw_polarization}. These modes, unlike the 
whistlers emerging as an extension of the fast mode, are essentially pressure balanced and carry 
significant density fluctuations.}. Note that while \eqref{eq:kaw_polarization0} and \eqref{eq:kaw_polarization} 
relate the Fourier coefficients of a single KAW, they do not strictly guarantee 
an analogous relation in real space for a linear combination of such waves. An exception to the latter
is the relation between $n_e$ and $b_\parallel$, which is given at any point in space
by Eq.~\eqref{eq:p_balance}. The (linearized) pressure balance is, however, a quite general
signature of low-frequency dynamics not exclusively limited to KAWs. For instance, 
it may include mirror structures \citep{Sahraoui2006,Pokhotelov2008,Genot2009,Perrone2016} 
or kinetic slow modes \citep{Narita2015}.
A less ambiguous method to identify KAW properties is 
to compare the relative spectral amplitudes between $|n_{e\vec k}|^2$ and $|\vec b_{\perp\vec k}|^2$
to the corresponding theoretical prediction, which is more exclusive to 
KAWs \citep{Chen2013}, and it is derived from the nonlinear equations 
by specifically assuming a (monochromatic) wave solution.
If pressure balance is taken for granted, the latter is equivalent to measuring 
the relative amplitudes between $|\vec b_{\parallel\vec k}|^2$ and $|\vec b_{\perp\vec k}|^2$.
While pressure balance alone
does not necessarily imply wave activity, it is still a fairly stringent condition at kinetic scales 
since it rules out high-frequency fluctuations such as 
whistler waves \citep{Galtier2003,Gary2009,Shaikh2009,Chen2017}, which cannot be \emph{a priori} neglected.
Moreover, the KAW and whistler wave predictions of $\bigl|\vec b_{\parallel\vec k}\bigr|^2/ 
\bigl|\vec b_{\perp\vec k}\bigr|^2$ may become degenerate for $\beta_i\gtrsim 1$ \citep{TenBarge2012a}.

It is straightforward to show \citep{Schekochihin2009} that a single monochromatic KAW is 
an exact solution of the fully nonlinear system \eqref{eq:ermhd1}-\eqref{eq:ermhd2}.
But that is not all. In analogy with shear Alfv\' en waves in MHD \citep{Boldyrev2006} 
and inertial waves in rotating fluids \citep{Nazarenko2011,Gelash2017}, a few special 
linear combinations of plane KAWs are as well exact solutions.
The exact wave solutions are generally found by requiring that the nonlinear terms vanish, 
which is satisfied whenever the contours of $n_e$, $\psi$, and $\nabla_\perp^2\psi$ are 
aligned in every perpendicular plane.
Unless all waves share the same $k_\perp$, the alignment between the contours of $\psi$ and $\nabla_\perp^2\psi$ 
restricts the geometry of the solutions. It then follows that the corresponding 
perpendicular profiles of $n_e$ and $\psi$ are either: (i) circularly symmetric or (ii) one-dimensional
(1D) [i.e., with a spatial variation along a single perpendicular direction]. In an unbounded domain, 
class (i) may be formally constructed as a Fourier integral of plane 
KAWs with cylindrically symmetric coefficients (with respect to $z$ axis), whereas (ii) corresponds to any 
combination of KAWs with a fixed direction of the 
perpendicular wave vector $\vec k_\perp$.
The exact wave solutions of type (i) and (ii) differ from the special class of waves 
with a fixed magnitude of $k_\perp$ found in Ref.~\citep{Schekochihin2009} in that they may be composed of 
counter-propagating plane KAWs. Moreover, the absence of a fixed $k_\perp$ constraint
permits wave packets with a spatially localized envelope. Strictly speaking, even if the 
perpendicular envelope of a KAW is spatially localized at a given time, it will not remain well localized at later times. 
However, since KAWs disperse primarily in the parallel direction, the perpendicular spreading of localized 
wave packets is relatively slow. In conclusion, the above aspects
suggest that KAWs are deeply rooted in the magnetized plasma dynamics, beyond the standard limits of 
a linear approximation.

The vanishing of nonlinear terms for solutions with either circularly symmetric or 1D 
perpendicular profiles is in fact a quite generic feature of turbulent wave
systems that possess a strong mean field, together with perpendicular nonlinearities in the form of a 
Poisson bracket: $\{f,g\} = \hat{\vec z}\cdot(\nabla_\perp f\times\nabla_\perp g)$.
This includes the strongly anisotropic ($k_\parallel\ll k_\perp$) limits of the 
MHD and rotating Navier-Stokes equations 
(e.g., see Ref.~\citep{Nazarenko2011}). For these two systems, the same types 
of exact linear wave solutions [namely (i) and (ii)] may be found in the ideal (i.e., dissipation-free) regime. For rotating
flows, the analogue of type (ii) solutions is derived explicitly in Ref.~\citep{Gelash2017}, 
even without assuming $k_\parallel\ll k_\perp$.
Moreover, class (i) relates to the monopole (force-free) Alfv\' en 
vortex solution in MHD \citep{Petviashvili1992, Alexandrova2008a}. Due to 
these similarities, some of our results might be 
directly relevant to structure formation in strongly anisotropic MHD and rotating turbulence.

\section{METHODS}
\label{sec:methods}

Below we describe the numerical simulation and observational data, the data analysis techniques, and introduce the generalized 
field ratios. These aspects are essential for a complete understanding of this work. However, those interested only 
in the main outcome of the study may skip over to the results in Sec.~\ref{sec:results}.

\subsection{Driven 3D fully kinetic simulation}
\label{sec:simulation}

We perform the simulation using 
the particle-in-cell code \textsc{osiris {\footnotesize 3.0}} \citep{Fonseca2002,Fonseca2013}. The periodic 
domain size is $(L_x, L_y, L_z) = (18.9,\,18.9,\,48.3)\,d_i$, where $d_i=\rho_i/\sqrt{\beta_i}$ is the ion inertial length. 
The global mean field $\vec B_0$ points in the $z$ direction. 
The spatial resolution is $(N_x,N_y,N_z)=(928,928,1920)$ with 150 particles per 
cell per species. Quadratic spline interpolation is used for the charge-conserving electric current
deposit. A reduced ion-electron mass ratio of $m_i/m_e=100$ is used and the electron plasma to cyclotron frequency 
ratio is $\omega_{pe}/\Omega_{ce} = 2.86$. To reduce particle noise, the output 
fields used for the analysis are short-time averaged
over a window of duration $\Delta t\Omega_{ce}= 2.0$, where $\Omega_{ce}$ is the electron cyclotron frequency.
Following Ref.~\citep{Zhdankin2017}, the forcing is an adaptation of the Langevin antenna \citep{TenBarge2014} and 
introduces a time-dependent external electric current $\vec J_{\rm ext}$. We apply $\vec J_{\rm ext}$ at wave numbers 
$(1,0,\pm1)$, $(0,1,\pm1)$, $(-1,0,\pm1)$, and $(0,-1,\pm1)$ in units of $(2\pi/L_x,\,2\pi/L_y,\,2\pi/L_z)$. The antenna 
current is divergence free and drives low-frequency Alfv\' enic fluctuations. To avoid 
a rapid transient response, we initialize the fluctuating field at $t=0$ to match the 
external current via $\nabla\times \vec b_{\perp} = (4\pi/c)\vec J_{\rm ext}$. We choose an antenna frequency 
$\omega_0 = 0.9\cdot (2\pi v_A/L_z)$, where $v_A=B_0/\sqrt{4\pi n_0 m_i}$ is the Alfv\' en speed, and the 
decorrelation rate \citep{TenBarge2014} is $\gamma_0 = 0.6\omega_0$. The ion and electron velocity distributions at $t=0$
are isotropic Maxwellians with $\beta_i\approx\beta_e\approx 0.5$.

\begin{figure}[htb!]
\includegraphics{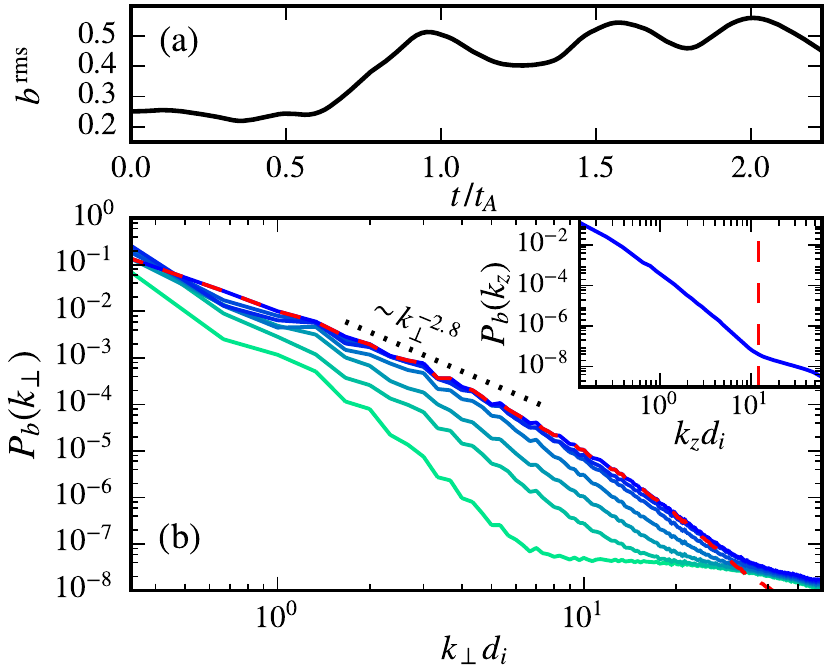}
\caption{Time trace of the root-mean-square fluctuating magnetic field (a) and the 1D $k_{\perp}$
magnetic spectra at times $t/t_A = \{0.66, 0.92, 1.18, 
1.45, 1.71, 1.97, 2.24\}$ in the simulation (light green to dark blue). A $-2.8$ slope is shown for reference (dotted line).
The inset shows the $k_z$ spectrum. Particle noise dominated modes with $k_zd_i>12$ (vertical dashed line) are 
filtered out in the subsequent analysis. The $k_z$ filtered $k_{\perp}$ spectrum at $t/t_A = 2.24$ is shown with a
red dashed line in panel (b). All components of the magnetic field
are normalized here to $B_0$.\label{pic:kaw_simulation}}
\end{figure}

The approach towards a statistically 
steady state is depicted in Fig.~\ref{pic:kaw_simulation}. The simulation is run for about 2.24 Alfv\' en transit 
times, $t_A = L_z/v_A$, until the kinetic-scale spectra appear to be converged. Towards the end of the simulation, the mean 
ion and electron betas (based on their space-averaged local values) are $\beta_i=0.56$ and $\beta_e=0.51$,
while the mean temperature anisotropy is $T_{\perp i}/T_{\parallel i}\approx 1.2$ for the ions
and $T_{\perp e}/T_{\parallel e}\approx 0.9$ for electrons.
The turbulence is strongly driven such that $b^{\rm rms} \approx L_{\perp}/L_z\approx 0.4$ towards 
the end of the simulation [Fig. ~\hyperref[pic:kaw_simulation]{\ref*{pic:kaw_simulation}(a)}], where $ b^{\rm rms}$ is 
the root-mean-square fluctuating field in units of $B_0$. This corresponds to
critical balance ($b^{\rm rms} \approx k_{\parallel}/k_{\perp}$) at the forced wave numbers.

\begin{figure}[htb!]
\includegraphics{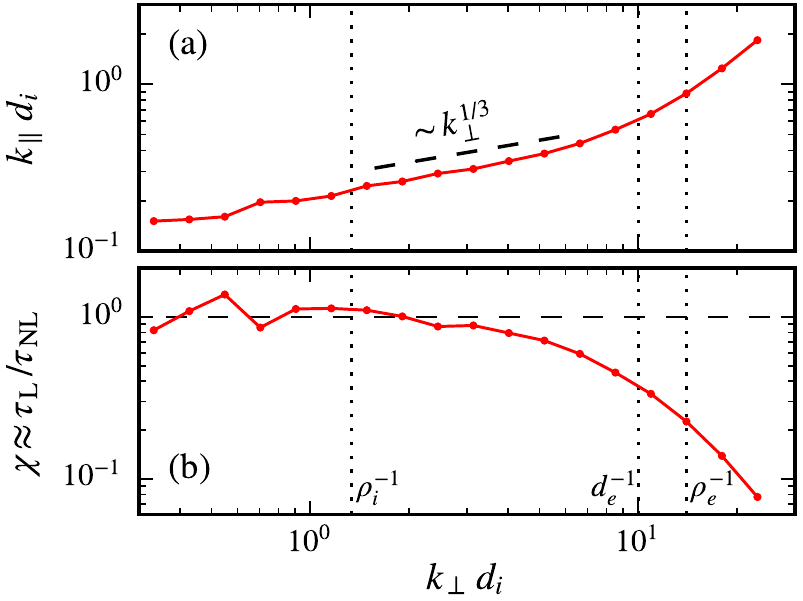}
\caption{\label{pic:anisotropy} Local anisotropy of the externally driven turbulence (a) and the ratio of 
linear to nonlinear time scales (b). A 1/3 slope in panel (a) is shown for reference.
All components of the magnetic field were normalized to $B_0$ in the calculation.}
\end{figure}

The obtained 1D magnetic energy spectrum as a function of $k_\perp\equiv(k_x^2+k_y^2)^{1/2}$
[Fig.~\hyperref[pic:kaw_simulation]{\ref*{pic:kaw_simulation}(b)}]
exhibits an approximate power law with a steepening of the spectral exponent close to electron scales. 
The spectrum is consistent with solar wind measurements, which typically show spectral exponents around $-2.8$ at sub-ion scales, while
steeper exponents are measured close to electron scales and beyond \citep{Alexandrova2009,Sahraoui2010}. In the inset of 
Fig.~\hyperref[pic:kaw_simulation]{\ref*{pic:kaw_simulation}(b)} we show for reference the 1D $k_z$ spectrum. 
To further reduce contributions from particle noise in the subsequent analysis, we filter out the noise dominated 
modes with $k_zd_i> 12$ \citep{Franci2018}. Employing the method of Cho \& \mbox{Lazarian} \citep{Cho2009}, 
we consider in Fig.~\ref{pic:anisotropy} 
the anisotropy relative to the \emph{local} mean field. The sub-ion range anisotropy, $k_\parallel(k_\perp)$, is 
scale-dependent and has a slope of approximately 1/3 on the logarithmic graph \footnote{Using an
alternative method based on a local, five-point structure function \citep{Cho2009} (not shown), we obtain
a somewhat weaker anisotropy with an approximate scaling $k_\parallel\sim k_\perp^{0.6}$.}, 
in agreement with a recent fully kinetic simulation of decaying turbulence \citep{Groselj2018}.
According to the 
asymptotic KAW theory for a $\beta\sim 1$ plasma \citep{Boldyrev2013}, KAWs are expected to exist for wave numbers 
with $k_{\parallel}/k_{\perp}\ll 1$ and $k_{\parallel}d_i\ll 1$. While these values are not asymptotically small
in our simulation, the estimated anisotropy is within the range 
$k_{\parallel}/k_{\perp}< 1$ and $k_{\parallel}d_i< 1$ at sub-ion scales.
We also estimate the ratio of linear (KAW) to nonlinear time scales 
[Fig.~\hyperref[pic:anisotropy]{\ref*{pic:anisotropy}(b)}],
given by $\chi = \tau_{\,\rm L} / \tau_{\,\rm NL} \approx \delta b_{\perp, k_{\perp}}k_{\perp} / k_{\parallel}$,
where $\tau_{\rm L}\approx 1/(k_\parallel k_\perp)$ and 
$\tau_{\rm NL}\approx 1/(k_\perp^2\delta b_{\perp,k_{\perp}})$ in dimensionless units 
\citep{Howes2008,Schekochihin2009,Sulem2016}.
To obtain the scale-dependent fluctuation $\delta b_{\perp,k_{\perp}}$,
we apply a band-pass filter between $k_\perp/2$ and $2k_\perp$ on $\vec b_\perp$ and compute its 
root-mean-square value \citep{Groselj2018}. Note that $\tau_{\rm L}$ and 
$\tau_{\rm NL}$ are based here on Eqs.~\eqref{eq:ermhd1}-\eqref{eq:ermhd2},
but their ratio $\chi$ takes the same form as the nonlinearity 
parameter in MHD if dynamic alignment is neglected \citep{Goldreich1995,Boldyrev2006,Mallet2015}. 
The ratio of linear to nonlinear 
time scales is close to unity for $k_{\perp}\rho_i\lesssim 1$ and exhibits a slight decline throughout the 
sub-ion range. This simple yet direct estimate of $\chi$ contrasts some previous 
works \citep{Matthaeus2014} and implies that the kinetic-scale nonlinear effects are not any more 
significant than linear wave physics, despite the fact that the turbulence fluctuation 
amplitudes are not small [Fig.~\hyperref[pic:kaw_simulation]{\ref*{pic:kaw_simulation}(a)}]. 
These circumstances provide additional motivation for comparison with linear KAW predictions.

In order to facilitate a direct comparison 
with observational data (see Sec.~\ref{sec:sw_data}), we construct a set 
of 1D traces from the simulation by mimicking a spacecraft fly-through with
several passings over the periodic box \citep{Greco2008}. A set of 100 virtual spacecraft
trajectories are analyzed to improve statistics. In particular,
we choose the direction $\hat{\vec n} = (0.949, 0.292, 0.122)$ and 
extract the fluctuations along this given direction using cubic spline
interpolation. Thus, the direction of extraction is quasi-perpendicular to $\vec B_0$.
The starting points of all linear trajectories are 
distributed uniformly in the $z=0$ plane. All trajectories end once they reach $z=L_z$.
To avoid spurious edge effects, we skip the wavelet coefficients 
that are within a distance (along the 1D trace) of $19 d_i$ from 
the edges of the trace when calculating the generalized field ratios (see Sec.~\ref{sec:wavelets}).

\subsection{Spacecraft data}
\label{sec:sw_data}

The solar wind data analysis is based on a 7 h interval from the Cluster spacecraft \citep{Escoubet2001} 
and on a 159 s interval from the Magnetospheric Multiscale (MMS) mission \citep{Burch2016}. 
These intervals were previously studied by Chen \emph{et al.}~\citep{Chen2015} and 
\mbox{Gershman} \emph{et al.}~\citep{Gershman2018}, respectively. At the time of the 
measurement, the Cluster spacecraft were in the free solar wind  far from the Earth's foreshock, whereas the 
MMS spacacraft were in the Earth's magnetosheath but well separated from the bow shock and the magnetopause. The mean plasma 
betas are $\beta_i\approx 0.26$ and $\beta_e\approx 0.62$ for Cluster and $\beta_i\approx 0.27$ and $\beta_e\approx 0.03$ for
the MMS interval. MMS spacecraft can also measure the mean temperature anisotropy, which was
$T_{\perp i}/T_{\parallel i}\approx 1.5$ for ions and $T_{\perp e}/T_{\parallel e} \approx 0.8$ for 
electrons during the interval.
The analyzed data includes magnetic measurements from Cluster \citep{Chen2015} and the electron density,
electron fluid velocity, and magnetic 
measurements from MMS \citep{Gershman2018}. We convert the spacecraft frame frequencies $f_{\rm sc}$
to field-perpendicular wave numbers using Taylor's hypothesis \citep{Howes2014}: 
$k_{\perp}\approx 2\pi f_{\rm sc}/v_0$, where $v_{0}$ is a characteristic velocity. We take $v_0$ to be the 
magnitude of the mean solar wind speed $\vec V_{\rm SW}$ for the Cluster interval. 
Given the relatively small angle between $\vec B_0$ and $\vec V_{\rm SW}$ during the MMS interval, 
we take $v_0 \approx |\vec V_{\rm SW}|\cos(\theta)$ for the MMS data, where $\theta\approx 76^\circ$ 
is the mean angle between $\vec V_{\rm SW}$ and the wave vector $\vec k$, determined in Ref.~\citep{Gershman2018} using the 
$k$-filtering technique. This angle was found to be relatively constant throughout the entire kinetic range. The inferred 
mean angle between $\vec k$ and $\vec B_0$, on the other hand, was very close to $90$ degrees \citep{Gershman2018}. To 
avoid edge effects, we skip a certain amount of points at the edges of the analyzed intervals when computing the wavelet
based diagnostics (see Sec.~\ref{sec:wavelets}). In particular, we skip about 5 s on each side of the Cluster interval
and about 7 s on each side of the MMS interval.

The MMS interval covers around 80 ion inertial lengths in the field-perpendicular direction. 
This is too short to allow for a statistically reliable analysis of intermittency and the results are 
included here for reference. Nevertheless, we still use the MMS data in order to be 
able to analyze simultaneous magnetic field and density measurements, which
is crucial for making direct contact with the KAW predictions, where density fluctuations play 
a major role \citep{Boldyrev2013,Chen2013}. Much longer suitable intervals from MMS are presently 
not available \citep{Gershman2018}. While the MMS interval is shorter than the typical large 
scale turbulence correlation time (e.g., \citep{Osman2014}), the Cluster interval features 
a relatively long continuous stream of 
measurements, covering several correlation times. It is thus more appropriate for studying intermittency, 
albeit with the limitation that only the magnetic measurements are available in this case. 

\subsection{Local scale extraction and generalized field ratios}
\label{sec:wavelets}

The turbulent fields are decomposed locally among scales using the complex-valued, 
continuous Morlet wavelet transform \citep{Farge1992,Torrence1998}. 
The 1D Morlet wavelet basis functions, $\psi_s^{\rm 1D}$, can be represented explicitly in spectral 
space as band-pass filters. Up to a normalization constant, the spectral representation is given by:
\begin{align}
\widehat{\psi}_s^{\,\rm 1D}(k)& = \Theta(k)\exp\left(\frac{-(k-k_s)^2}{2(k_s/k_\psi)^2}\right),\label{eq:morlet1d}
\end{align}
where $\Theta(k)=1$ for $k>0$ and 0 otherwise, 
$k_\psi= 6$ is a dimensionless parameter \citep{Farge1992} and
$k_s$ is a characteristic wave number scale. The variable $k_s$ 
is related to the wavelet scale $s$ via $k_s = k_\psi/s$.
For some field $f(x)$, 
the Morlet wavelet coefficients $\widetilde{f}_s(x)$ at scale $s$ are obtained from the inverse 
Fourier transform of $\widehat f\,\widehat\psi_s$, where $\widehat f$ is the Fourier transform of $f$. The set of 
wave number scales $\{k_s\}$ is logarithmically spaced. We define a local, scale-dependent fluctuation as 
$\delta f_s(x) = c_1 \Re\bigl\{ \widetilde{f}_s\bigr\}$, where $\Re\{\dots\}$ is the
real part. Similarly, we define a local power spectral density 
as $P_f(k_s, x) = c_2 \bigl|\widetilde{f}_s\bigr|^2/k_s$. The normalization
constants $c_1$ and $c_2$ may be determined based on the exact parameters of the wavelet transform
\citep{Farge1992,Torrence1998}. We set $c_1=c_2=1$ since our results do 
not dependent on such constant prefactors. In the following, we drop the scale subscript $s$ 
but it is to be understood that all quantities are scale dependent. 

When comparing the simulation results with spacecraft measurements, the 1D Morlet 
wavelet transform is applied to a set of virtual spacecraft trajectories extracted from the 
simulation (see Sec.~\ref{sec:simulation}). In Fig.~\ref{pic:slices}, we also employ 
a 2D generalization of the Morlet transform \citep{Farge1992,Kirby2005}.
The corresponding basis functions are represented in spectral space as
 \begin{align}
\widehat{\psi}_{s,\phi}^{\,\rm 2D}(k_x, k_y)& = \exp\left(\frac{-(k_x-k_s\cos\phi)^2}{2(k_s/k_\psi)^2}\right)&\nonumber\\
&\;\;\;\,\times\exp\left(\frac{-(k_y-k_s\sin\phi)^2}{2(k_s/k_\psi)^2}\right),\label{eq:morlet2d}&
\end{align}
where $\phi$ determines the direction of extraction, $k_\psi=6$, and $k_s$ is defined same as above.
We also impose $\widehat{\psi}_{s,\phi}^{\,\rm 2D}(0, 0)=0$. Twelve directions for $\phi$ in the 
range from $0$ to $\pi$ are used and the final results are angle-averaged.
The angular averaging of the local spectrum is performed after taking
the squared modulus of the wavelet coefficient \citep{Kirby2005}.

We use the wavelet decomposition to define a new set of statistical measures. In particular, we 
consider the spectral field ratios, frequently used to study wave properties 
(e.g., \citep{Gary2009,Salem2012,Chen2013,Cerri2017,Groselj2018}),
and generalize their definitions to investigate the 
impact of large-amplitude, turbulent structures on these ratios. Two sets of generalized ratios are defined. 
The first set is based on the scale-dependent moments \citep{Farge2015} of the fluctuations:
\begin{align}
\left(\frac{\left\langle|\delta n_e|^m\right\rangle}{\left\langle|\delta b_{\perp}|^m\right\rangle}\right)^{\!2/m},&&
\left(\frac{\left\langle|\delta n_e|^m\right\rangle}{\left\langle|\delta b_{\parallel}|^m\right\rangle}\right)^{\!2/m},&&
\left(\frac{\left\langle|\delta b_{\parallel}|^m\right\rangle}{\left\langle|\delta b_{\perp}|^m\right\rangle}\right)^{\!2/m},
\label{eq:moments}
\end{align}
where $m$ is the order of the moment and $\langle\dots\rangle$ represents a space average. In the simulation,
we additionally average over a set of 100 virtual spacecraft trajectories (see Sec.~\ref{sec:simulation}).
The parallel and perpendicular components are defined here 
relative to the \emph{local} 
mean field $\vec B_{\rm loc}$ as
$\widetilde b_{\parallel} = \widetilde{\vec b}\cdot\hat{\vec b}_{\rm loc}$ and
$\widetilde{\vec b}_{\perp} = \widetilde{\vec b} - \widetilde b_{\parallel} \hat{\vec b}_{\rm loc}$,
where $\hat{\vec b}_{\rm loc} = \vec B_{\rm loc} / |\vec B_{\rm loc}|$ \citep{TenBarge2012a,Kiyani2013}. 
The local mean field is obtained from a Gaussian low-pass filter with a standard 
deviation $\sigma_s  = k_s/k_\psi$ in spectral space. 
Once the parallel and perpendicular components are determined, 
we normalize the fluctuations according to \eqref{eq:norm_den}-\eqref{eq:norm_bpar}.
For $m=2$ the moment ratios yield the standard spectral ratios, defined in terms of the 1D $k_\perp$ 
spectra. On the other hand, for $m>2$ the averaging becomes progressively more sensitive to the 
fluctuations at the tails of the probability distribution function, thus giving insight into the
dependence of the ratios on high-amplitude events. We consider moments up to $m=6$.

Caution is needed when computing high-order 
statistics from finite data sets, since the tails of the probability distribution function may not be 
sufficiently sampled \citep{DudokdeWit2004,Kiyani2006}.
For example, an estimate of the maximal moment $m_{\max}$ that can be determined accurately, 
based on the method presented in Ref.~\citep{DudokdeWit2004}, yields typical values of $m_{\max}$ between 3 and 4 for the Cluster 
interval with $N\approx 6.3\times 10^5$ samples \footnote{Estimates of $m_{\max}$ for MMS are 
scattered between 1 and 6 at different scales. If the sample size is relatively small, 
the estimate of $m_{\max}$ itself might be rather inaccurate.}. 
To obtain more reliable estimates, we employ the scheme of Kiyani 
\emph{et al.}~\citep{Kiyani2006} (see also Ref.~\citep{Chen2014}) and remove a small fraction of the largest fluctuations at each scale
until the moments appear reasonably converged.
For Cluster 
we find that removing 0.005\% (i.e., about 30 samples) of largest fluctuations  
seems to be adequate. 
For consistency, we clip the same small fraction in the simulation when calculating the moments.
Within the statistical uncertainties, the clipping does not significantly affect the results 
and only makes it easier to recognize true statistical trends.
Due to the short duration of the MMS interval, the 0.005\% clipping threshold has no effect and 
relatively large fractions would have to be removed 
to make the moments well behaved. Thus, no attempt is made to recover 
more reliable estimates via clipping for the MMS interval.

The ratios introduced in \eqref{eq:moments} are global measures in a sense that the average is 
taken over the entire ensemble. A more local 
measure can be obtained via conditional averaging of the local power spectral densities:
\begin{align}
\frac{\langle|\widetilde n_e|^2\,|{\rm LIM}>\xi\rangle}{\langle|\widetilde b_{\perp}|^2\,|{\rm LIM}>\xi\rangle},&&
\frac{\langle|\widetilde n_e|^2\,|{\rm LIM}>\xi\rangle}{\langle|\widetilde b_{\parallel}|^2\,|{\rm LIM}>\xi\rangle},&&
\frac{\langle|\widetilde b_{\parallel}|^2\,|{\rm LIM}>\xi\rangle}{\langle|\widetilde b_{\perp}|^2\,|{\rm LIM}>\xi\rangle},
\label{eq:lims}
\end{align}
where $\rm{LIM}$ is the local intermittency measure \cite{Farge1992,RuppertFelsot2009} 
and $\xi$ is the threshold for the conditional average. We scan a range of different thresholds and 
study how the results depend on this choice.
The $\rm LIM$ is defined as the local wavelet spectrum normalized to 
its mean at a given scale. Thus, $\rm LIM>1$ at the locations where the power spectral density exceeds its mean value.
The $\rm LIM$ may be based on different quantities. We use everywhere the 
same type of $\rm LIM$ so that all conditional averages are constrained  
to the same spatial locations. 
In particular, we choose the $\rm LIM$ based on the KAW energy density (see Sec.~\ref{sec:theory}): ${\rm LIM}  \equiv \bigl(|\widetilde b_{\perp}|^2 
+ |\widetilde n_{e}|^2\bigr)/ \bigl\langle |\widetilde b_{\perp}|^2 + |\widetilde n_{e}|^2\bigr\rangle$. This 
appears to be a reasonable choice given that the kinetic-scale structures carry both significant 
magnetic field and density fluctuations as shown in what follows. For Cluster measurements, $|\widetilde b_\parallel|^2$ 
is used as a proxy for $|\widetilde n_e|^2$ to obtain the $\rm LIM$ 
under the assumption of pressure balance \eqref{eq:p_balance}.
The conditional averages may eventually become energetically insignificant, since the 
averaging volume shrinks with the threshold $\xi$. To focus on the conditional ratios which
are still of some energetic relevance, we require for any averaging sub-domain to contain at least 1\% of the total KAW energy at that 
scale. Estimates not satisfying the condition are omitted from the results. The typical 
volume fractions corresponding to the 1\% energy 
fraction are naturally much smaller than 1\%.

The generalized ratios \eqref{eq:moments}-\eqref{eq:lims} are 
evaluated from turbulence data and
compared against the linear predictions \eqref{eq:kaw_polarization}. 
In the normalized, beta-dependent
units \eqref{eq:norm_den}-\eqref{eq:norm_bpar}, linear KAW theory predicts a 
numerical value of unity for all ratios considered above in the asymptotic limit \eqref{eq:asymptotic_limit}. 
Since the asymptotic range is vanishingly small in the simulation, we also compare our
results against the linear wave predictions obtained from a fully kinetic 
numerical dispersion solver \citep{Verscharen2018}.
Note that the KAW predictions are formally obtained for a single plane wave. 
An ensemble of KAW packets with different propagation directions 
may exhibit slight statistical deviations from the
expected values for those ratios in \eqref{eq:moments}-\eqref{eq:lims}
that depend on $\vec b_\perp$ (i.e., for the ratios that 
are not based on the pair of fields $n_e$ and $b_\parallel$, which are directly related
via pressure balance; see Sec.~\ref{sec:theory}).

\section{RESULTS}
\label{sec:results}

\begin{figure}[htb!]
\includegraphics{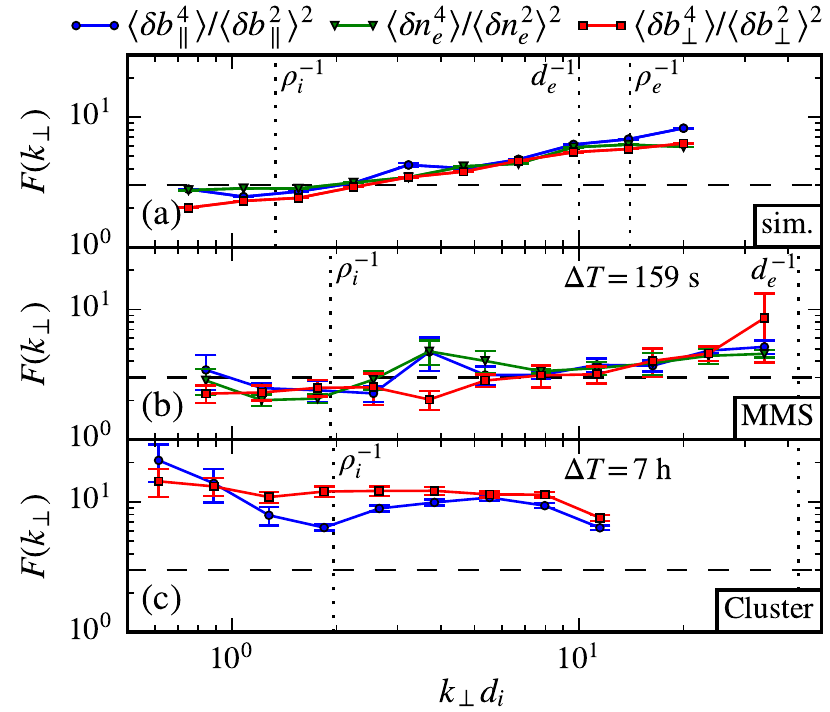}
\caption{Scale-dependent flatness results obtained from the simulation (a), 
MMS interval (b), and the Cluster interval (c). Vertical dotted lines mark the plasma kinetic scales and the 
horizontal dashed lines mark the Gaussian value of 3.\label{pic:flatness}}
\end{figure}

We now turn to the main results of this work. First, we characterize the statistical nature of fluctuations separately for each field 
in terms of the scale-dependent flatness \citep{Farge2015}: $F(k_{\perp}) = \langle|\delta f|^4\rangle / \langle|\delta f|^2\rangle^2$, where 
$\delta f\in\{\delta n_e, \delta b_{\parallel}, \delta b_\perp\}$ represents a wavelet decomposed field (sec Sec.~\ref{sec:wavelets}). 
For a field with a Gaussian probability distribution, we have $F=3$. Thus, high values of the flatness above 3 
characterize the degree of non-Gaussianity, while a scale-dependence of $F$ points towards intermittency in the classical sense 
of the term as a departure from self-similarity \citep{Frisch1995}. 
The scale-dependent flatness results obtained from solar wind measurements 
and from the 3D fully kinetic simulation are compared in Fig.~\ref{pic:flatness}. 
To illustrate 
the statistical uncertainties, we add errorbars to the flatness measurements.
To obtain the errorbars, 
we calculate the moments separately on a number of non-overlapping 
subsets and then use these as input for a jackknife error estimate of the flatness \citep{Shao1995}. 

As evident from the 
results presented in Fig.~\ref{pic:flatness}, the kinetic-scale 
fluctuations exhibit signatures of non-Gaussian statistics
with flatness values above 3. The departure from Gaussian
statistics is altogether largest for the Cluster interval. 
In agreement with previous works \citep{Kiyani2009, Kiyani2013}, the Cluster statistics are nearly scale-independent 
at kinetic scales. This is in contrast with the simulation where the flatness is scale-dependent and only gradually 
increases above the Gaussian value with decreasing scale, presumably due to the finite simulation 
domain size. All three analyzed fields in the simulation exhibit similar statistical properties, in agreement
with the Cluster measurements of $\delta b_\perp$ and $\delta b_\parallel$ below the $\rho_i$ scale. 
It is worth pointing out that a previous analysis of 
density intermittency in the free solar wind \citep{Chen2014} found $\delta n_e$ flatness values 
comparable to our Cluster results, which do not include the electron density.
As already mentioned (Sec.~\ref{sec:sw_data}), the MMS interval is too short to allow for a 
statistically reliable analysis of intermittency. Nevertheless, within the relatively large uncertainties, 
the turbulence statistics appear to be mildly intermittent for the MMS interval.

\begin{figure}[htb!]
\includegraphics{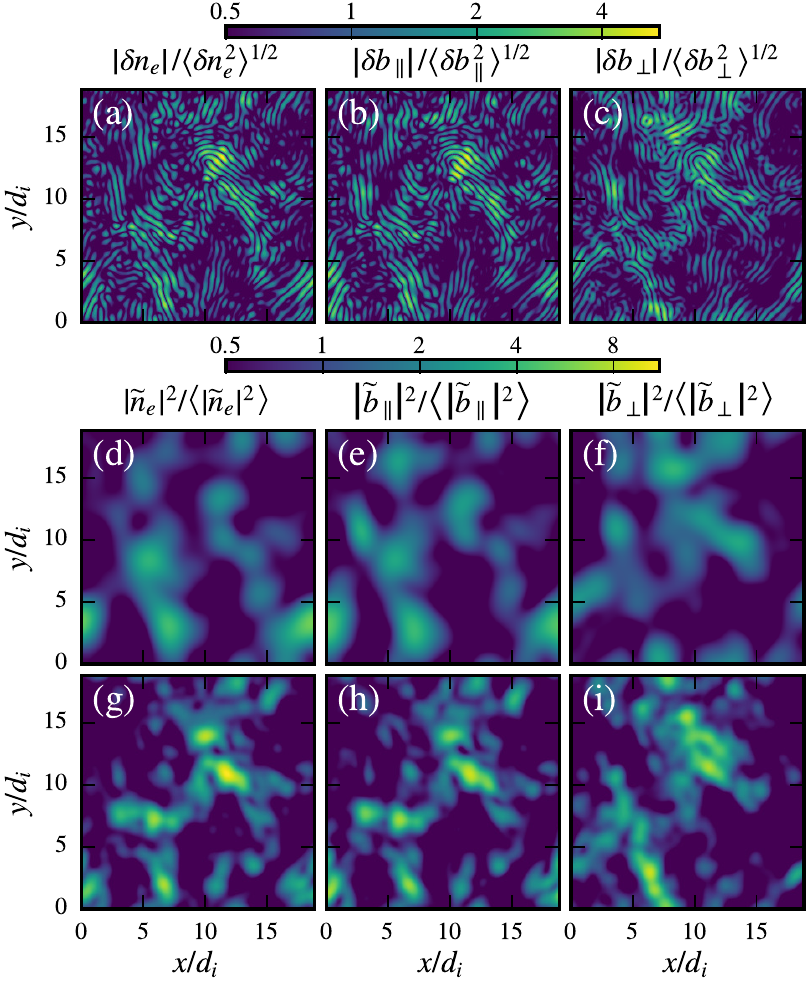}
\caption{Scale-filtered fluctuations in a given $x$--$y$ plane in the range $k_\perp d_i=[5, 10]$ (a)--(c),
and the (normalized) local wavelet spectra at scale  $k_\perp d_i=5$ (d)--(f) and at scale $k_\perp d_i=10$ (g)--(i).
A logarithmic scale is used to better show also the fluctuations of moderate intensity. Very weak fluctuations with
amplitudes below 0.5 in the normalized units are clipped to the lower boundary of the color map.
\label{pic:slices}}
\end{figure}

Next, we inspect the spatial structure of fluctuations in the simulation. The 
fluctuations in a given $x$--$y$ plane are visualized in Fig.~\ref{pic:slices}. 
In Figs.~\hyperref[pic:slices]{\ref*{pic:slices}(a)--\ref*{pic:slices}(c)}
we plot the fields in the range $k_\perp d_i=[5, 10]$ by summing up the wavelet decomposed fluctuations 
in that range \citep{Farge1992,He2012,Roberts2013,Perrone2016} using 8 logarithmically spaced scales. 
Note that $k_\perp d_i=10$ already corresponds to the electron inertial scale $d_e=0.1d_i$ 
due to the reduced ion-electron mass ratio employed here.
A distinct feature seen in Fig.~\ref{pic:slices} is the 
excellent matching of the $\delta n_e$ and $\delta b_\parallel$ fluctuation profiles. Although not obvious 
considering the full range of kinetic effects retained in the simulation, the latter directly
implies that the structures are pressure balanced according to Eq.~\eqref{eq:p_balance}. 
In Figs.~\hyperref[pic:slices]{\ref*{pic:slices}(d)--\ref*{pic:slices}(i)}
we consider the local 2D wavelet spectra 
at $k_\perp d_i=5$ [Figs.~\hyperref[pic:slices]{\ref*{pic:slices}(d)--\ref*{pic:slices}(f)}]
and at  $k_\perp d_i=10$ [Figs.~\hyperref[pic:slices]{\ref*{pic:slices}(g)--\ref*{pic:slices}(i)}]. The spatial distribution of the spectral 
energy density is non-uniform and the peaks in the spectra at different scales tend to be concentrated 
around the same spatial locations. This local coupling across different scales is a characteristic
feature of coherent structures \citep{Farge1992, RuppertFelsot2009, Lion2016}.
It is also seen that the non-uniformity increases at smaller scales, 
consistent with the growth of the flatness with $k_\perp$ (Fig.~\ref{pic:flatness}). Finally, while the 
$n_e$ and $\vec b_\parallel$ wavelet spectra match very well, the local $\vec b_\perp$ spectra match 
the former only in a rather loose sense. This is as well consistent with theoretical expectations, 
since no general linear relation exists between $\vec b_\perp$ and $n_e$ in real space
according to the KAW turbulence theory (see Sec.~\ref{sec:theory}). The fact that an apparent 
weak coupling is seen at all points towards 
the importance of nonlinear effects in shaping the local fluctuations.

\begin{figure}[htb!]
\includegraphics[]{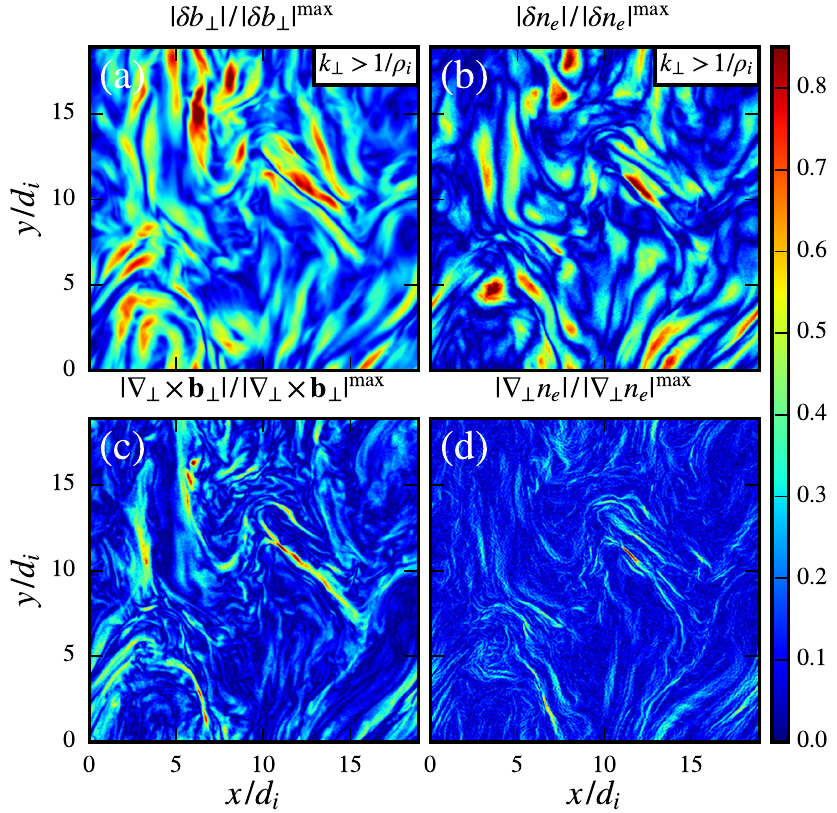}
\caption{Turbulent structures in a perpendicular simulation plane as seen from 
the perpendicular magnetic (a) and density fluctuations (b), and from their corresponding
perpendicular gradients (c)--(d). Large-scale modes with $k_\perp \leq 1/\rho_i$, dominated 
by external forcing, have been filtered out in (a) and (b) to highlight 
the sub-ion scale structure.}\label{pic:jz_wz_contours}
\end{figure}

Since aggressively scale-filtered results can be a somewhat misleading, we show in Fig.~\ref{pic:jz_wz_contours}
the perpendicular magnetic and density fluctuations over the entire sub-ion range and beyond, together
with their corresponding (non-filtered) perpendicular gradients. The perpendicular cross-sections of the
various sub-ion scale structures broadly resemble either sheets or circular shapes.
A 3D inspection of the fluctuations (not shown) confirms that the 
structures are indeed elongated in the field-parallel direction, 
consistent with the anisotropy estimate presented in Fig.~\ref{pic:anisotropy}.
It is also interesting to notice, that the small-scale perpendicular gradients of 
$\vec b_\perp$ and $n_e$ tend to form structures that are aligned with respect to 
each other (see also Sec.~\ref{sec:discussion}).

\begin{figure}[htb!]
\includegraphics{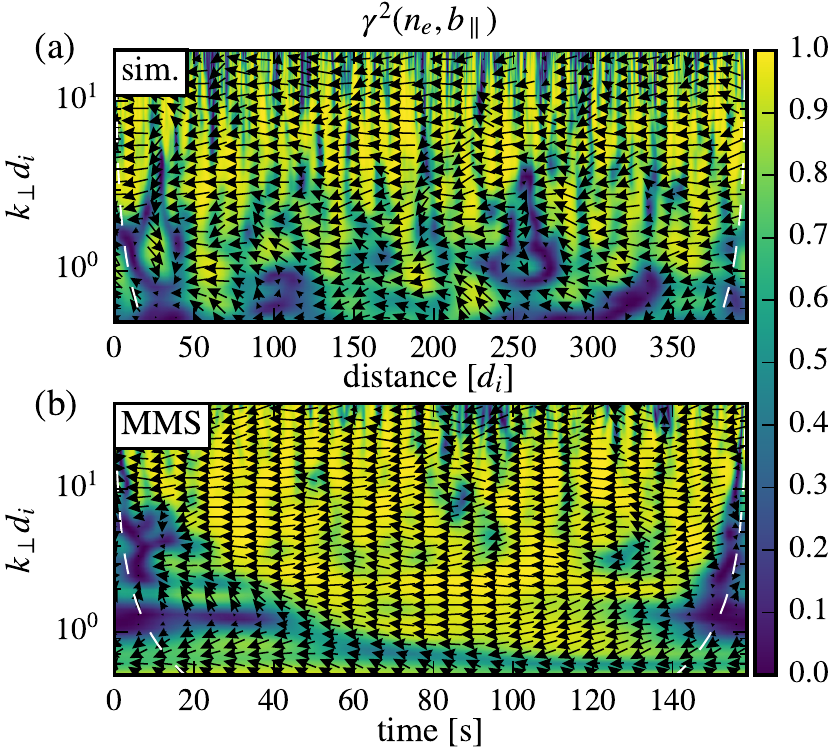}
\caption{Magnitude squared cross-coherence and phase between $n_e$ and $b_\parallel$ in the simulation (a) and in the 
MMS interval (b). The orientation of the 
arrows (relative to the positive horizontal axis) denotes the phase. Dashed lines show the cone of influence \citep{Torrence1998}.
\label{pic:coherence}}
\end{figure}

To further investigate the perpendicular pressure balance \eqref{eq:p_balance}, we 
compute the wavelet cross-coherence \citep{Grinsted2004, Lion2016, Chen2017}
between $n_e$ and $b_\parallel$ in the simulation and for the MMS interval, using the 
1D Morlet wavelet transform \citep{Torrence1998}. High values of cross-coherence close to unity indicate a strong local phase 
synchronization between two signals. The results are compared in Fig.~\ref{pic:coherence}. 
Arrows are used to show the phase between the two fields. 
A strong phase synchronization is seen at sub-ion scales of the simulation and 
of the MMS interval. With most arrows pointing to the left in Fig.~\ref{pic:coherence}, the results strongly
suggest that the density and parallel magnetic field fluctuations tend to be anti-correlated and thus fulfill
the pressure balance \eqref{eq:p_balance} to a good approximation \footnote{A few events, 
mostly around the $d_i$ scale, are seen in the simulation where either the cross-coherence is 
low or where the scale-dependent phase between $n_e$ and $b_\parallel$ 
significantly differs from 180 degrees. A possible reason could be that the fluctuations 
(especially at large scales) are generally too large to allow 
for a linearization of the magnetic pressure \citep{Schekochihin2009} around its local background value.}. 
This conclusion is in agreement with previous works based on MMS data \citep{Chen2017,Gershman2018}.
Qualitatively similar results were also obtained by studies of pressure balance in the
MHD-scale range (e.g., \citep{Howes2012a,Klein2012}). As noted in Sec.~\ref{sec:theory}, 
pressure balance is a rather generic feature of 
low-frequency dynamics and as such it rules out the high-frequency whistler waves.
We mention that we also checked the wavelet coherence between
$n_e$ and (different components of) $\vec b_\perp$. As expected (see Sec.~\ref{sec:theory}), 
$n_e$ and $\vec b_\perp$ generally do not exhibit a strong cross-coherence 
(not shown). However, at rare times of high coherence we often observe a relative 
phase close to 90 degrees, consistent with the 
elliptical polarization of the KAW \citep{Schekochihin2009,He2012}.

\begin{figure*}[htb!]
\includegraphics{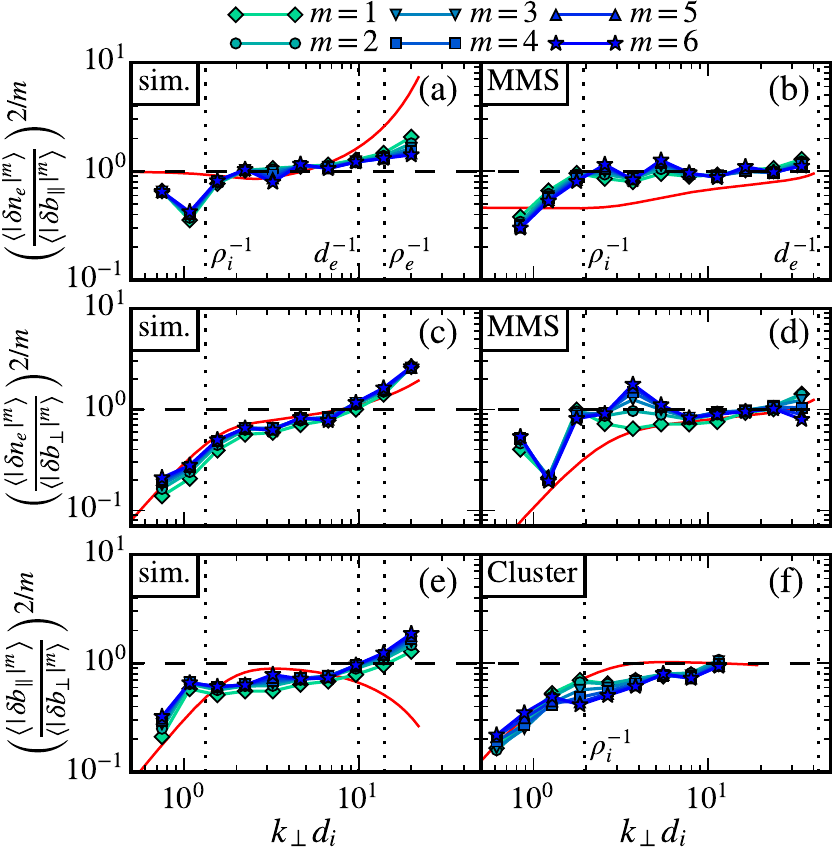}
\includegraphics{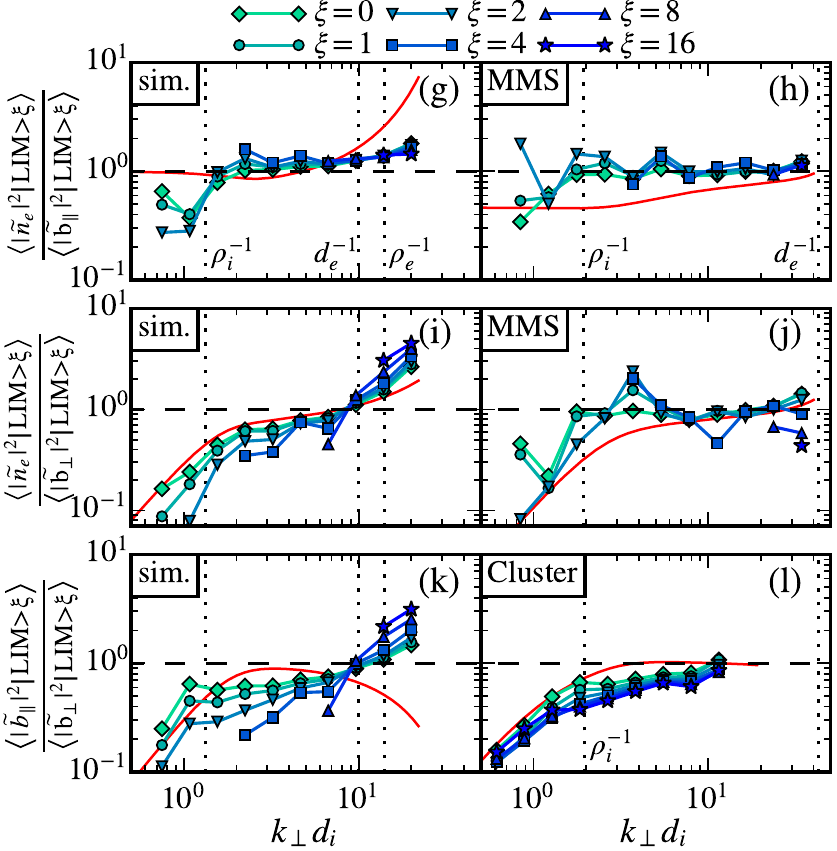}
\caption{Generalized spectral field ratios obtained from Cluster and MMS measurements 
and from the 3D fully kinetic simulation (see text for further details). Panels (a)--(f) show the 
moment ratios \eqref{eq:moments} and panels (g)--(l) show the conditional ratios \eqref{eq:lims}.\label{pic:ratios}}
\end{figure*}

Finally, we turn to the central subject of the present paper and present the generalized field ratios results.
The generalized ratios obtained from the 3D fully kinetic simulation and from spacecraft measurements are 
plotted in Fig.~\ref{pic:ratios}. 
In Figs.~\hyperref[pic:ratios]{\ref*{pic:ratios}(a)--\ref*{pic:ratios}(f)} we show the 
moment ratios \eqref{eq:moments} and in Figs.~\hyperref[pic:ratios]{\ref*{pic:ratios}(g)--\ref*{pic:ratios}(l)} we display 
the conditional ratios \eqref{eq:lims}, conditioned on the local KAW energy density (see Secs.~\ref{sec:theory} and \ref{sec:wavelets}). 
Dashed horizontal lines denote the 
linear asymptotic KAW predictions \eqref{eq:kaw_polarization}, while red lines show the more accurate 
linear predictions obtained from a fully kinetic plasma dispersion relation solver \citep{Verscharen2018}. 
A wave propagation angle of 89.9 degrees (with respect to $\vec B_0$) was used for solving 
the numerical dispersion relations \footnote{Over the range of scales where 
ion cyclotron resonance is absent, the 
predicted field ratios are nearly independent of the exact choice 
of the highly oblique wave propagation angle.
However, for angles slightly below the chosen value of 89.9$^\circ$, the 
numerically obtained dispersion relations
exhibit cyclotron resonance before reaching electron scales. 
Due to the distinct nonlinear frequency broadening in strong turbulence (e.g., \citep{Meyrand2016}), 
it is reasonable to expect that the turbulent fluctuations are able
to cascade beyond the resonance gap to frequencies $\omega>\Omega_{ci}$, where 
coupling to the KAW branch is again possible (see Refs.~\citep{Howes2008,Boldyrev2013,Lopez2017}).}. 
Dotted vertical lines indicate the characteristic ion and electron
scales. The choice $m=2$ or $\xi=0$ for the moment ratios and
conditional ratios, respectively, yields the standard spectral ratios.
We mention that the maximal meaningful threshold $\xi$ (see Sec.~\ref{sec:wavelets}) grows with $k_\perp$ 
for the simulation and for MMS data since the 
fluctuations become more intermittent at smaller scales. The Cluster interval exhibits highly non-Gaussian 
statistics for all $k_\perp$ (Fig.~\ref{pic:flatness}) and thus allows for the use of high thresholds over the whole range.
The higher the moment $m$ or threshold $\xi$, the more sensitive the ratios are to the large-amplitude events. 
Note that the term ``large-amplitude'' refers here to the locally enhanced 
fluctuations and power spectral densities, compared to their scale-dependent mean square values. 
Spatially, these intense bursts of local activity are not distributed incoherently but instead form 
distinct patterns (Figs.~\ref{pic:slices} and \ref{pic:jz_wz_contours}), typically associated with turbulent coherent structures. The
locally enhanced, non-Gaussian fluctuations go hand in hand with the non-uniform power spectral densities, since it can be shown that
the spatial variability of the energy spectrum and the scale-dependent flatness are directly related \citep{Bos2007}.
Higher moments assign smaller statistical weights to the lower-amplitude background fluctuations in favor of the large-amplitude
events. Similarly, the conditional averages discard the locations 
with a spectral energy density below the threshold $\xi$ and therefore measure
the field ratios within the most energetic structures only. 

As seen in Fig.~\ref{pic:ratios},
the generalized ratios either follow the linear wave predictions (red lines) throughout 
the transition region ($k_\perp \sim 1/\rho_i$),
where shear Alfv\' en waves convert into KAWs, or converge toward the KAW predictions below the
$\rho_i$ scale. This is consistent with KAW theory, where $\rho_i$ is the relevant transition 
scale \citep{Schekochihin2009, Boldyrev2013, Kunz2018}. A notable deviation from linear wave 
predictions is seen at sub-electron scales of the simulation. A similar trend was previously 
seen in gyrokinetic simulations \citep{TenBarge2012a,Toldcomm}, which possibly implies that the
wavelike activity is terminated below the $\rho_e$ scale via 
collisionless damping \citep{TenBarge2013a}. The moment ratios depend only weakly on the
order $m$, which shows that the wave predictions are reasonably satisfied not only in terms of
low-order statistics but also in the sense of higher-order moments related to intermittency.
The conditional ratios between $|\widetilde n_e|^2$ and $|\widetilde b_\parallel|^2$
[Figs.~\hyperref[pic:ratios]{\ref*{pic:ratios}(g)--\ref*{pic:ratios}(h)}] 
appear to be somewhat less scattered compared to the other two conditional ratios 
[Figs.~\hyperref[pic:ratios]{\ref*{pic:ratios}(i)--\ref*{pic:ratios}(l)}], which exhibit a slight tendency to
deviate further from linear predictions with increasing LIM threshold $\xi$.
On the other hand, the sub-ion scale deviations attributed to large-amplitude 
events are altogether only moderate, such that the generalized ratios remain in reasonable agreement 
with the KAW polarization properties. 
Thus, the wavelike features are not exclusively limited to low-amplitude 
fluctuations but also carry over to the high-amplitude structures. Qualitatively similar 
results are obtained in the simulation as well as from spacecraft measurements, which suggests 
that the observed properties are a common feature of the kinetic turbulence.
While some of the measured ratios might allow for alternative interpretations (see Sec.~\ref{sec:theory}), 
the fact that an agreement is seen for \emph{all} generalized ratios and for different values 
of the ambient ion and electron plasma beta (recall that the normalizations 
\eqref{eq:norm_den}-\eqref{eq:norm_bpar} are beta-dependent) implies
that the fluctuations are indeed of the KAW type. Moreover, the hypothetical possibility that the 
field polarizations are not related to wavelike activity is contradicted by 
our estimate of the nonlinearity
parameter (Fig.~\ref{pic:anisotropy}), and we are unaware of any other concrete 
theoretical prediction that could consistently explain all of our measurements.
The observed order unity preservation of linear wavelike properties in kinetic-scale 
turbulent structures constitutes the main result of this work, together with the 
supplemental evidence presented in Figs.~\ref{pic:anisotropy}--\ref{pic:coherence}.

\section{DISCUSSION}
\label{sec:discussion}

The general idea that intermittent turbulent structures may go hand in hand with wave physics was implied 
in a number of recent works.
In a numerical study of MHD turbulence \citep{Mallet2015}, a scale-independent probability distribution 
with an order unity mean was found for the 
local (in space) ratio between the linear-wave and nonlinear time scale,
whereas all other quantities of interest exhibited scale-dependent, intermittent statistics. 
The former property was later successfully exploited to design a statistical model of intermittency
in MHD turbulence \citep{Mallet2017a}. Moreover, Howes~\citep{Howes2016a} recently developed a dynamical 
model for current sheet formation via nonlinear Alfv\' en wave collisions. 
Subsequent numerical studies
showed that current sheets emerge also in a more realistic
configuration with two counter-propagating, initially separated wave packets
\citep{Verniero2018}, 
and that the modes produced during nonlinear wave interaction can be themselves characterized as 
Alfv\' en waves \citep{Verniero2018a}. The above-described developments 
in astrophysical plasma physics are not a unicum in turbulence research. For instance, similar ideas have been 
put forward in the context of rotating, inertial wave turbulence \citep{Staplehurst2008}.

In KAW turbulence, wave packet interactions are relatively complex 
due to the dispersive nature of the KAW which allows for the nonlinear coupling between
co-propagating waves \citep{Schekochihin2009,Voitenko2016}. This prevents a direct 
application of the model by Howes~\citep{Howes2016a}, which is based on incompressible MHD equations, 
to sub-ion scale dynamics. On the other hand, our results shown in Fig.~\ref{pic:ratios}
imply that the sub-ion scale turbulent 
structures may as well be viewed as nonlinearly 
interacting wave packets. This quantitative evidence encourages the design of new 
models of turbulent heating in low-collisionality plasmas that incorporate intermittency
by exploiting the wavelike character of coherent structures \citep{Mallet2018}. Additional evidence is also provided
by a recent observational study \citep{Chen2019} and by gyrokinetic turbulence 
simulations \citep{TenBarge2013,Klein2017},  which find the energy transfer
between fields and particles to be dominated by particles satisfying the resonance condition
for Landau damping of (kinetic) Alfv\' en waves, even though the transfer 
tends to be spatially concentrated 
in the vicinity of intense current sheets \citep{Wu2013a,Osman2014,Wan2015,Howes2018}.

Since the nonlinear time of a turbulent eddy is inversely proportional
to its fluctuation amplitude \citep{Matthaeus2014}, one might be tempted to conclude that the
large-amplitude nonlinear structures evolve on a time scale faster than the wave dynamics.
A number of aspects may be pointed out here. Unlike the nonlinear time, the linear wave time 
is proportional to $1/k_\parallel$ \citep{Howes2008,Schekochihin2009}.
For a realistic turbulent setup, the latter may be interpreted 
as a parallel correlation length $\ell_\parallel\sim 1/k_\parallel$ 
of an eddy \citep{Schekochihin2009, Sulem2016}.
In magnetized plasma turbulence, information is transmitted along the field lines by waves.
It then follows from a causality principle that the maximal parallel correlation 
is set by the distance a wave packet can cover within the
lifetime of an eddy \citep{Boldyrev2005,Schekochihin2009,Nazarenko2011}. 
This implies that the wave time scale is approximately limited from above by the nonlinear time, 
which is indeed consistent with our measurement of the anisotropy 
in Fig.~\ref{pic:anisotropy}. An analogous causality 
argument can be given for rotating turbulence in the strongly 
anisotropic ($k_\parallel\ll k_\perp$) regime \citep{Nazarenko2011}.
It was also recently argued, based on a set of 3D gyrokinetic simulations \citep{Verniero2018a}, 
that the non-propagating $k_\parallel=0$ modes do not play a notable role in the mediation 
of energy transfer in MHD, provided that appropriate measures are 
taken to avoid periodic boundary artifacts.
Secondly, it is worth considering the empirical fact that turbulence does not 
self-organize into structures of arbitrary geometry. Instead, a number of studies of kinetic range
turbulence find that the structures 
tend to acquire relatively simple shapes, resembling either 
elongated sheets or tubes \citep{Smith2011,Boldyrev2012,TenBarge2013,
Wan2015,Perrone2016,Kobayashi2017,Wang2019}.
For $k_\parallel \neq 0$, the idealized geometric versions of these two (i.e., sheets with a 1D perpendicular profile 
and circularly symmetric, field-aligned tubes) correspond to exact 
wave solutions of the nonlinear equations \eqref{eq:ermhd1}-\eqref{eq:ermhd2}, 
as explained in Sec.~\ref{sec:theory}. 
The wandering of magnetic field lines in turbulent flows precludes such highly symmetric configurations \citep{Boldyrev2006,Eyink2013,Howes2017}. However, even if a structure 
only resembles an idealized sheet or circular tube, the nonlinearity is locally weakened.
Some of these circumstances were also appreciated by Terry \& Smith \citep{Terry2007, Smith2011}, 
who argued that the magnetic shear at the edges of large-amplitude circular filaments and sheets 
prevents the structures from mixing with the background incoherent KAW fluctuations. 
Moreover, local depletions of the nonlinearity 
have been associated with the emergence of coherent structures in a variety of 
turbulent systems (e.g., \citep{Farge2001,Servidio2008,Matthaeus2008a,Bos2008,Pushkarev2014}).

\begin{figure}[htb!]
\includegraphics{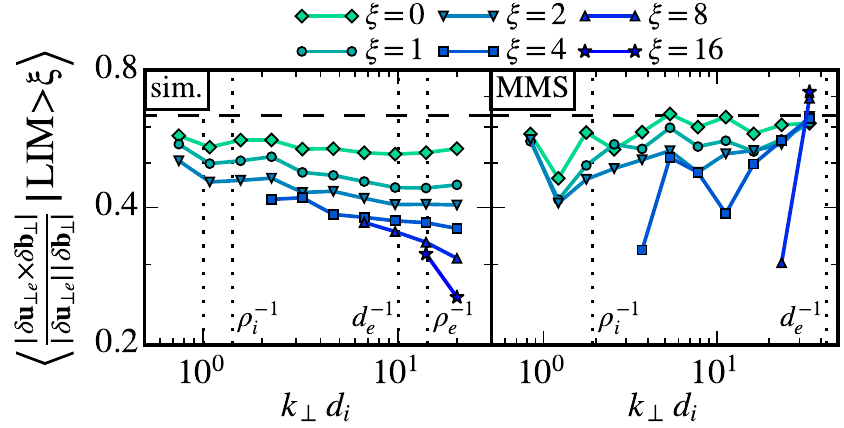}
\caption{Sine of the scale-dependent alignment angle between the perpendicular electron
fluid velocity and magnetic field, conditionally averaged on a given LIM threshold 
$\xi$ (see text for details). Dashed horizontal lines indicate the 
expected value for a randomly distributed angle.
\label{pic:alignment}}
\end{figure}

Is it possible to provide evidence for the above suggestion? To this end, we
make use of the fact that in the first approximation $\nabla_\perp n_e\propto \hat{\vec z}\times\vec u_{\perp e}$ 
and $\nabla_\perp\psi\propto \hat{\vec z} \times \vec b_\perp$ at sub-ion scales \footnote{See 
Appendix C of Ref.~\citep{Schekochihin2009} for details. In the simulation, we checked that 
the small-scale structure of $\hat{\vec z}\times\vec u_{\perp e}$ is indeed very similar 
to the field structure of $\nabla_\perp n_e$. Note that high-resolution measurements of 
$\vec u_{\perp e}$ are readily available from MMS, while the density gradient is much more 
challenging to estimate.}, where $\vec u_{\perp e}$ is the perpendicular electron fluid velocity.
If $\vec u_{\perp e}\times \vec b_\perp = 0$, the nonlinear coupling in Eq.~\eqref{eq:ermhd1} vanishes.
We then consider the scale-dependent sine of the alignment angle, 
$\sin\theta\equiv |\delta\vec u_{\perp e}\times \delta\vec b_{\perp} |/|\delta\vec u_{\perp e}||\delta\vec b_\perp|$, 
conditionally averaged on a given LIM threshold
for the MMS and simulation data. Same as in the rest of this paper, the scale decomposition is 
implemented using the Morlet wavelet transform and the perpendicular components are defined 
with respect to the local mean field. Our results are shown in Fig.~\ref{pic:alignment}.
According to the above discussion, the alignment angle should be
preferentially reduced within the energetic structures (corresponding to high LIM values).
The results in Fig.~\ref{pic:alignment} are indeed consistent with our expectation, 
although the preferential alignment within energetic structures is less pronounced for the MMS interval. 
The latter may be related to the fact that the fluctuations in the MMS interval are only 
mildly intermittent (see Fig.~\ref{pic:flatness}). We also add that an intermittent field
alignment, analogous to the one studied here over the kinetic range of scales, was 
previously reported in the context of MHD turbulence (e.g., \citep{Beresnyak2006,Mallet2016}).
Here, we suggest that such intermittent, kinetic-scale alignment helps to preserve 
the linear KAW properties in large-amplitude structures via local nonlinearity depletions.

Finally, there exist numerous other examples of the importance of waves in 
complex nonlinear plasma phenomena. Classic examples include 
the nonlinear steepening of MHD waves into 
shocks (e.g., \citep{Cohen1974,Hada1985,Neugebauer1990,Mann1995,Lee2000}) and 
solitary waves in magnetized plasmas \citep{Hasegawa1976,Petviashvili1992,
Spangler1982,Shukla1985,Wu2004}. More recently, there has been a growing
interest to relate soliton solutions with kinetic-scale structures in 
turbulent space plasmas \citep{Sundkvist2005,Lion2016,Perrone2016,Wang2019}.
Indeed, it cannot be ruled out that at least some turbulent structures are 
better described as solitons than dynamically 
evolving wave packets. Similar to the exact linear wave solutions 
discussed in Sec.~\ref{sec:theory}, soliton solutions
typically require a rather idealized, highly symmetric and/or low-dimensional geometry.
Thus, for the same reason as the turbulent structures cannot be exactly
linear waves (namely, stochastic field line wandering), they are probably not precisely 
solitons either, unless perhaps if the system is far from a state of homogeneous, 
fully developed strong turbulence.

\section{CONCLUSIONS}
\label{sec:conclusions}

In the present paper, we employ high-resolution observational and kinetic simulation data to
study the relationship between wavelike physics and structure formation in astrophysical kinetic plasma turbulence.
Observational data are based on \emph{in situ} solar wind measurements from the Cluster and MMS spacecraft, and the simulation
results are obtained from an externally driven, 3D fully kinetic simulation.
We introduce appropriate generalizations of the frequently employed spectral field ratios to probe the 
statistical field polarizations within the large-amplitude energetic structures. We find that both the lower
amplitude background fluctuations as well as the intense structures satisfy linear KAW predictions to
order unity. This implies that the turbulent structures \emph{themselves} approximately preserve 
a linear wave footprint. In other words, our novel analysis highlights the possibility that wavelike
features and turbulent structure formation are essentially inseparable from each other.
As such, our results challenge one of the presently
common views of waves and structures ``coexistence.'' Furthermore, it is suggested that 
a linear wave character in kinetic-scale structures is preserved with the aid of local 
nonlinearity depletions.
 
The only known framework capable of providing a reasonable theoretical basis 
for the interpretation of our results appears to be the KAW turbulence phenomenology 
\citep{Howes2008,Schekochihin2009,Boldyrev2013}. Known alternatives, such as whistler turbulence \citep{Galtier2003,Gary2009,
Shaikh2009,Gary2012}, cannot explain 
our results. In contrast to our results, the high-frequency whistlers are not pressure balanced and they
carry only minor density fluctuations \citep{Chen2013,Chen2017}. In a $\beta\sim 1$ plasma, whistler waves are also 
expected to be rather strongly damped for $k_\parallel d_i\lesssim 1$ \citep{Boldyrev2013}, 
which contradicts our anisotropy estimate in Fig.~\ref{pic:anisotropy}.
In the KAW turbulence context, an interpretation similar to recent developments 
in MHD turbulence emerges (e.g., \citep{Howes2016a,Mallet2017a,Verniero2018,Verniero2018a}). Namely, the 
kinetic-scale structures could be perhaps described as nonlinearly evolving, 
localized KAW packets. 
The possibility that at least a fraction of the 
sub-ion scale structures are sheetlike, as implied by our simulation results (see also Refs.~\citep{Boldyrev2012,Kobayashi2017}), 
lends credence to recent works emphasizing the role of reconnection 
in sub-ion scale turbulence \cite{Cerri2017a,Mallet2017,Loureiro2017,Franci2017,Phan2018}.
In our present understanding, the results of this work do not preclude the reconnection scenario, 
although the mere presence of sheetlike structures alone is not a sufficient condition for 
a reconnection-mediated type of turbulence.

Finally, we mention that the general approach employed 
here is not exclusively limited to kinetic range turbulence in astrophysical plasmas and we hope it might find 
exciting applications in a broad range of turbulent systems where waves and structures have been observed 
\citep{Benkadda1994,Smolyakov2000,Lindborg2006,Staplehurst2008,Zhdankin2013,Godeferd2015,Biferale2016}. 
An immediate extension of the method lends itself in the context of MHD range plasma turbulence, where 
it could be used to study the interplay between structures and waves based on a 
generalized Alfv\' en ratio \citep{Salem2009,Chen2016}. Another potential application includes
rotating turbulence \citep{Staplehurst2008,Nazarenko2011,Godeferd2015,Biferale2016}, 
where the generalized spectral field ratios could be employed to test the
presence of inertial-wave polarization locally within the 
energetic columnar structures.

\begin{acknowledgments}

We gratefully acknowledge helpful conversations with D. \mbox{Told}, A. \mbox{Ba{\~ n}{\' o}n} \mbox{Navarro}, 
J.~M. \mbox{TenBarge}, S.~S. \mbox{Cerri},  B.~D.~G. \mbox{Chandran}, and 
A.~A. \mbox{Schekochihin}. D.G. thanks in particular to 
N.~F. \mbox{Loureiro} for the fruitful exchange of ideas related to
this work and for his assistance with obtaining the computing resources,
and F.~\mbox{Tsung}, V.~\mbox{Decyk}, and W.~\mbox{Mori} for discussions 
on particle-in-cell methods and simulations with the \textsc{osiris} code. 
C.H.K.C. was supported by a Science and Technology Facilities Council (STFC) 
Ernest Rutherford Fellowship No.~ST/N003748/2, and A.M. by National Science Foundation
(NSF) Grant No.~AGS-1624501. K.S. acknowledges support by the French Research Federation for Fusion 
Studies carried out within the framework of the European Fusion 
Development Agreement (EFDA).
The Cray XC40, Shaheen,  at the King Abdullah University of Science \&
Technology (KAUST) in Thuwal, Saudi Arabia was utilized for all the
simulations reported.
The authors would like to acknowledge the OSIRIS Consortium, consisting of 
UCLA and IST (Lisbon, Portugal) for the use of \textsc{osiris {\footnotesize 3.0}} and for providing 
access to the \textsc{osiris {\footnotesize 3.0}} framework. 

\end{acknowledgments}

%\bibliography{refs_aps}

%%%%%%%%%%%%%%%%%%%%%%%%%%%%%%%%%%%%%%%%%%%%%%%%%%%%%%%%%%%%%%%%%%%%%%%%%%%%%%%%%%%%%%%%%%%%%%%%%%%%%%%%


%%%%%%%%%%%%%%%%%%%%%%%%%%%%%%%%%%%%%%%%%%%%%%%%%%%%%%%%%%%%%%%%%%%%%%%%%%%%%%%%%%%%%%%%%%%%%%%%%%%%%%%%

\end{document}